\documentstyle[aps,prb,%%%%%%
multicol,bezier%%%%%%%%%<--------- REMOVE BEFORE SUBMISSION
]{revtex}
\begin{document}
\title{
   Inhomogeneous states in quantum dots. I.
}
\author{ Daniel L. Miller }
\address{ Dept. of Physics of Complex Systems,\\
          The Weizmann Institute of science,
          Rehovot, 76100 Israel                \\
          e-mail  fndaniil@wicc.weizmann.ac.il
}
\date{\today}
\maketitle
\begin{abstract}
   In this work we show that the Anderson impurity model applied to
  ``scar'' wave function may explain large fluctuations of
   ground state energy of electron gas in a quantum dot.
\end{abstract}
\begin{multicols}{2}
\narrowtext

Single electron wave functions $\psi_j(\vec r)$ of classically chaotic
system are inhomogeneous.  Consider the unstable periodic orbit $p$ in a
system with two degrees of freedom. Such an orbit contributes to the density
of states at the energy $E$ the sum\cite{Gutzwiller-mar71}
\begin{equation}%%%%%%%%%%%%%%%%%%%%%%%%%%%%%%%%%%%%%%%%%%%%%%%%%%%%%%%
    {- T_p\over 2\pi\hbar}   \text{ Im }
    \sum_{m=1}^\infty
    {1\over \sinh(\lambda_p m/2)}
    e^{im S_p(E+i0)/\hbar} \;,
\label{eq:trace.1}
\end{equation}
where $T_p={\partial S_p/\partial E}$ is the period of the orbit, $S_p(E)$
is the action of the orbit, $S_p(E)$ includes the phase changes at the
conjugated points, $\lambda_p>0$ is the stability exponent.

Let us assume that $p$ is the only one relatively stable
$\lambda_p \lesssim 1 $
periodic orbit in the system. In this particular case $p$ contributes
a sequence of sharp Lorentzians\cite{Gutzwiller-mar71}
\begin{equation}%%%%%%%%%%%%%%%%%%%%%%%%%%%%%%%%%%%%%%%%%%%%%%%%%%%%%%%
    \sum_n
    T_p(E) { \lambda_p \hbar \over
   [ S_p(E) - 2\pi n \hbar ]^2 + (\lambda_p \hbar /2)^2 }\;.
\label{eq:trace.3}
\end{equation}
Each term in the sum can be regarded as the spectral function of the ``scar''
state
\begin{equation}%%%%%%%%%%%%%%%%%%%%%%%%%%%%%%%%%%%%%%%%%%%%%%%%%%%%%%%
   A_n(E) =
   {1\over \pi} {\Gamma_n \over ( E -  E_n)^2 + \Gamma_n^2}\;,
\label{eq:decau.1}
\end{equation}
valid under the condition
\begin{equation}%%%%%%%%%%%%%%%%%%%%%%%%%%%%%%%%%%%%%%%%%%%%%%%%%%%%%%%
   {\Gamma_n\over E_{n+1}-E_n}
   \approx {\lambda_p\over 4\pi} \ll 1\;,
\label{eq:trace.2}
\end{equation}
since $S_p(E_n) = 2\pi n \hbar$ and $\Gamma_n=\lambda_p \hbar /(2T_p(E_n))$.

This spectral function tell us that it is possible to construct
non-stationary solution $\psi_n(r)$ to the Schroedinger equation in the
vicinity of $p$. This solution, let us call it
``scar''\cite{Heller-Tomsovic-93},  would decay to ``flat'' states with the
rate $\Gamma_n/\hbar$. The matrix elements of the decay process $V_{jn}$ are
not known, but it is clear that
\begin{equation}%%%%%%%%%%%%%%%%%%%%%%%%%%%%%%%%%%%%%%%%%%%%%%%%%%%%%%%
    \Gamma_n = \pi \sum_j |V_{jn}|^2 A_{n}(\varepsilon_j)\;,
\label{eq:decay.5}
\end{equation}
where $\varepsilon_j$ are energies of the exact wave functions. Fortunately
we will not need explicit values of $V_{jn}$ in further calculations.

In what follows let us assume that the system is filled by electrons
at zero temperature. Let $n=d$ is the first ``scar'' state below the Fermi
energy $E_F$.  The Hamiltonian for the states in the vicinity of the Fermi
level is
\begin{eqnarray}%%%%%%%%%%%%%%%%%%%%%%%%%%%%%%%%%%%%%%%%%%%%%%%%%%%%%%%
   \hat {\cal H} &=&  \sum_\sigma\biggl\{
      \sum_{j, \varepsilon_j > E_d-W_d}^{\varepsilon_j < E_d+W_d }
      \left[ \varepsilon_j a_{j\sigma}^\dagger a_{j\sigma}
      + V_{jd} a_{j\sigma}^\dagger a_{d\sigma} + \text{h.c.}
   \right]
\nonumber\\
   &+ & E_d a_{d\sigma}^\dagger a_{d\sigma}\biggr\}\;,
\label{eq:decay.6a}
\\
   W_d &=& (E_{d+1}-E_d) /2\;,
\label{eq:decay.6}
\end{eqnarray}
and the formulation of the Anderson impurity model\cite{Anderson-may61} is
accomplished by adding the correlation energy
\begin{equation}%%%%%%%%%%%%%%%%%%%%%%%%%%%%%%%%%%%%%%%%%%%%%%%%%%%%%%%
   \hat{\cal H}_{\text{corr}} = U \hat n_{d\uparrow}\hat n_{d\downarrow}
\;,
\label{eq:decay.7}
\end{equation}
with
\begin{equation}%%%%%%%%%%%%%%%%%%%%%%%%%%%%%%%%%%%%%%%%%%%%%%%%%%%%%%%
   U = \int d\vec r d\vec r'
   |\psi_{d}(\vec r)|^2 |\psi_{d}(\vec r')|^2
   {e^2\over |\vec r - \vec r'|}
   - U_0\;.
\label{eq:decay.8}
\end{equation}
Here $U_0 \sim e^2/L$ is the charging energy of the system,  $L$ is the
typical size of the system.

Let us consider experiment, where one fills the system by particles. The
Fermi level goes up $E_F(N)\equiv E(N)-E(N-1)\propto N$, where $N$ is the
number of particles inside the system, and $E(N)$ is the energy of the ground
state of $N$ particles.  At the value $N=N_d$ given by equation
\begin{equation}%%%%%%%%%%%%%%%%%%%%%%%%%%%%%%%%%%%%%%%%%%%%%%%%%%%%%%%
   E_F(N_d)  =  E_d + U/2
\label{eq:model.1}
\end{equation}
(assuming $U\lesssim W_d$)
we arrive at the symmetric Anderson model. The author have
shown\cite{Miller-com98} that just below this value of $N$ the Fermi energy
makes a large jump
\begin{equation}%%%%%%%%%%%%%%%%%%%%%%%%%%%%%%%%%%%%%%%%%%%%%%%%%%%%%%%
   E_F(N_d) - E_F(N_d-1) \equiv \Delta_\ast\sim
   (U\Gamma \Delta)^{1/3}
\label{eq:model.2}
\end{equation}
that is just a large positive fluctuation of the level spacing. Here $\Delta$
is  the mean level spacing,  further increase of the Fermi level should be
``smooth'',  $\delta E_F\approx \delta N \Delta $. The negative fluctuation
is also possible and may be understood as relaxation of the pseudogap.

The Fermi energy will jump
again near $N=N_{d+1}$. The distance between jumps is $ N_{d+1}-N_d\approx
(E_{d+1}-E_d) / \Delta $ is inverse proportional to the length of
the orbit $l_p$
\begin{equation}%%%%%%%%%%%%%%%%%%%%%%%%%%%%%%%%%%%%%%%%%%%%%%%%%%%%%%%
   N_{n+1} - N_{n} \approx {2L\over l_p}
   \sqrt{2\pi N_{n+1}}\;.
\label{eq:model.3}
\end{equation}
This formula was derived for $\Delta \approx \pi \hbar^2/(mL^2)$ and it is
valid for large $N$, $N_{n+1} - N_{n} \ll N_{n+1}$.

The second derivative of the ground state  energy
$\Delta_2(N)=E(N+1)+E(N-1)-2E(N)$ was measured in the experiment of
Sivan {\em et al}\cite{Sivan-aug96}; it shows the sequence of particularly
strong fluctuations. They may be explained by the present theory with
$l_p\sim 2L$.   The numerical results of Stopa\cite{Stopa-sep97} supports the
present idea.  For example, the observed levitation of ``scar'' state near
the Fermi level is the signature of the Anderson impurity model.

The contribution of $p$ to the Green function of the non-interacting
system has the form $G(r,r)\propto e^{iW(x)y^2/\hbar}$, where $x$ is the
coordinate along $p$, $y$ axis is perpendicular to the orbit $p$ at the point
$x$, and $W(x)$ is connected with the second derivatives of the action.
Then, the width of the ``scar'' state is estimated as $\sim \sqrt{\hbar/W}$.
It is clear that $W(x)\sim k/L$, where $k=\sqrt{2mE_n}/\hbar$, and $m$ is the 
mass of the particle, see for example the analysis of chaotic 
billiards\cite{Bogomolny-88}.  The integral Eq.~(\ref{eq:decay.8}) for the 
wave function concentrated in the rectangle of the width $\sim\sqrt{L/k}$ and 
length $L$ gives $U\sim {e^2\over L} (\log{\sqrt{k L}}-1)$. For experiment of 
Sivan {\em et al}\cite{Sivan-aug96} $k L\sim 100$ and the correlation energy 
$U$ is large enough to make $\Delta_\ast$ noticeable.

To summarize, inhomogeneities of chaotic wave functions can be gathered
together into additional state of small size. In this way one arrives at the
Anderson's impurity model with finite number of free electron states. At
certain value of the chemical potential the model has complete particle-hole
symmetry. The chemical potential does not approach this value
gradually, but rather irregularly, see our companion paper. The energy scale 
of this effect is a new combination of the parameters of  Anderson's model
and the mean level spacing. This scenario is a possible explanation of the 
irregularities observed experimentally in quantum dots.

\acknowledgments
This work was supported by Israel Science Foundation and the Minerva Center
for Nonlinear Physics of Complex systems.

\end{multicols}

\begin{references}

\bibitem{Gutzwiller-mar71}
M.~C.~Gutzwiller, J. Math. Phys. {\bf 12},  343  (1971).

\bibitem{Heller-Tomsovic-93}
E.~J.~Heller and S.~Tomsovic, Phys. Today {\bf 46},  38  (July 1993).

\bibitem{Anderson-may61}
P.~W.~Anderson, Phys.~Rev. {\bf 124},  41  (1961).

\bibitem{Miller-com98}
Daniel~L.~Miller,  [Preprint cond-mat/9808089].

\bibitem{Sivan-aug96}
U.~Sivan, R.~Berkovits, Y.~Aloni, O.~Prus, A.~Auerbach, and G.~Ben-Yoseph,
  Phys. Rev. Lett. {\bf 77},  1123  (1996).

\bibitem{Stopa-sep97}
M.~Stopa,  [preprint cond-mat/9709119].

\bibitem{Bogomolny-88}
E.~B.~Bogomolny, Physica D {\bf 31},  169  (1988).

\end{references}
\end{document}